\documentclass
{elsarticle}

\usepackage{natbib}

\usepackage{amssymb}
\usepackage{graphicx}

\begin{document}

\begin{frontmatter}



\title{Adsorption of molecular gases on porous materials in the SAFT-VR approximation}


%

\author[label1]{M. Castro}
\author[label1]{R. Martinez}
\author[label2]{A. Martinez}
\author[label1]{H.C. Rosu}\ead{hcr@ipicyt.edu.mx}

\address[label1]{IPICyT, Instituto Potosino de Investigacion Cientifica y Tecnologica, \\ Apartado Postal 3-74 Tangamanga, 78231 San Luis Potos\'{\i}, Mexico}
\address[label2]{Divisi\'on de Ciencias e Ingenier\'{\i}as, Campus Le\'on, Universidad de Guanajuato, \\ Lomas del Bosque 103, Colonia Lomas del Campestre, Le\'on 37150, M\'exico}

\begin{abstract}
A simple molecular thermodynamic approach is applied
to the study of the adsorption of gases of chain molecules on solid surfaces. We use a
model based on the Statistical Associating Fluid Theory for Variable Range (SAFT-VR) potentials
[A. Gil-Villegas, A. Galindo, P. J. Whitehead, S. J. Mills,
G. Jackson, A. N. Burgess, J. Chem. Phys. 106 (1997) 4168] that we extend by including a quasi-two-dimensional approximation to
describe the adsorption properties of this type of real gases [A. Mart\'{\i}nez, M. Castro, C. McCabe, A. Gil-Villegas, J. Chem. Phys. 126 (2007) 074707].
The model is applied to ethane, ethylene, propane, and carbon dioxide adsorbed on activated carbon and silica gel, which are porous media of significant industrial interest. We show that the adsorption isotherms obtained by means of the present SAFT-VR modeling are in fair agreement with the experimental results provided in the literature.
\end{abstract}

\begin{keyword}
adsorption \sep statistical associating fluid \sep Helmholtz free energy \sep porous material
\PACS 68.43.-h \sep 51.30.+i

\bigskip
\bigskip

elsarticle-Martin1.tex \hfill Physica A 389, 3140-3148 (2010)\\
\hfill doi:10.1016/j.physa.2010.04.028


\end{keyword}

%

\end{frontmatter}



\newpage

\section{Introduction}

The main component of natural gas is methane, but
small amounts of other gases such as ethane,
ethylene, propane, propadiene, butane, methylpropane, 2-methylpropane, and carbon dioxide
are present. These components are
in negligible amounts in natural gas but become
a problem when the methane is stored
through adsorption and are less volatile than the methane and therefore preferentially occupy the active sites of
the adsorbent. To solve the problem of this type of pollutants
one must know their specific adsorbate properties on
various adsorbents at ambient temperatures and a broad range of
pressures.

The study of adsorption is very important for designing a
good adsorbent. A correct interpretation and quantification of the
adsorption isotherms are
required for the development of adsorption technologies
in the case of the separation and purification of gases and liquids
\cite{Vansant 1998,Tarasevich 1998}. A detailed  molecular thermodynamic model
based on perturbation theory for fluids
with highly anisotropic interactions has been developed by Wertheim \cite{Wertheim 1984} some time ago and has been known
as the Statistical Associating Fluid Theory (SAFT) since the works of Chapman and collaborators \cite{chapman 1989,chapman 1990}.

The SAFT approach in three dimensions has been used in a variety of systems and
complexes relevant to the industry, and due to its predictive power
and versatility started to be widely used in chemical engineering.
Recently, Martinez and collaborators \cite{martinez 2007} modified the two dimensional form of SAFT-VR approach (SAFT-VR-2D) to a quasi two-dimensional variant and obtained very good results in the case of adsorbed
gases such as methane and  nitrogen on activated carbon and in addition for as complex molecules as
asphaltene  adsorbed on some natural porous rocks present in an oil well \cite{castro 2009}.

This paper presents the application of the SAFT-VR formalism,
both in the context of 2D \cite{martinez 2007,castro 2009,sagrario 2008} and 3D \cite{gil 1997}, to obtain the adsorption isotherms
of ethane ($C_{2}H_{4}$), ethylene ($C_{2}H_{6}$), propane ($C_{3}H_{8}$), and
carbon dioxide ($CO_{2}$) in the case of activated carbon and silica gel as adsorbents at pressures up to 4 MPa and for three different temperatures.
\section{Theory}
In this section, we summarize the principal terms that enter the SAFT-VR model in 2D and 3D and the way we combine them to get the quasi two-dimensional approximation.

\subsection{Adsorption of monomeric fluids}\label{sub1}

The model consists of a simple fluid of $N$ particles with diameter $\sigma$ in the presence of a
uniform wall \cite{martinez 2007, castro 2009}. The interaction exerted by the wall on the particles is of the square well (SW) type
%
\begin{equation}
u_{pw}(z)=\left\{
\begin{array}{ll}
\infty & \mbox{$z<0$}\\
-\epsilon_{w} & \mbox{$0<z<\lambda_{w}\sigma$}\\
0 & \mbox{$\lambda_{w}\sigma<z$}~,\\
\end{array}
\right.
\end{equation}
where $z$ is the perpendicular distance of the particles
from the wall, $\epsilon_{w}$ is the depth of the well, and
$\lambda_{w}\sigma$ is the range of the attractive potential.

The system is divided into two subsystems: a fluid whose
molecules are near the wall forming the adsorbed fluid considered as a substrate of the wall and a fluid
whose molecules are far from the wall forming the fluid in the bulk. The fluids,
both the adsorbed and the bulk phases, have different properties
because of the presence of the wall. In particular, it is well known that the interaction
between the molecules depends on how close they are to the wall, in other words
the binary interaction between particles is different in the adsorbed and the bulk phases \cite{pitzer 1960}.
Thus, the particle density in the fluid, $\rho$, is a function
of $z$. To describe the amount of adsorbed particles, we define the following parameter
\begin{equation}\label{e2}
\Gamma=\int_{0}^{\infty}[\rho(z)
-\rho_{b}]dz~,
\end{equation}
where $\rho_{b}$ is the particle density in the bulk region. Since the length scale
in the adsorbed fluid is defined by $\lambda_{w}\sigma$, we can rewrite Eq.~(\ref{e2}) as
\begin{equation}\label{e3}
\Gamma=\int_{0}^{\lambda_{w}\sigma}
\rho(z)dz-\rho_{b}\lambda_{w}\sigma
\end{equation}
where the integral in the right hand side of Eq.~(\ref{e3}) is the density of the adsorbed particles, $\rho_{ads}$.
In thermodynamic equilibrium, the chemical potential of the adsorbed phase $\mu_{ads}$ and the bulk phase $\mu_{b}$ should be
equal.
Therefore, $\rho_{ads}$ can be obtained from the condition
\begin{equation}\label{e5}
\mu_{ads}=\mu_{b}~.
\end{equation}

The partition function of the adsorbed fluid is given by
\begin{equation}\label{e6}
Z_{ads}=\frac{V_{ads}^{N}}{N!\Lambda^{3N}}Q_{ads}~,
\end{equation}
where $V_{ads}^{N}$ is the volume containing the adsorbed fluid, $\Lambda$ is the de Broglie thermal wavelength in terms of
Planck's constant, and $Q_{ads}$ is the configurational partition function of the adsorbed system.

Introducing the adsorption area $S$, we have $V_{ads}= S\lambda_{w}\sigma$, and $Q_{ads}$ can
be rewritten as \cite{martinez 2007,castro 2009,sagrario 2008}
\begin{equation}\label{e7}
Q_{ads}=Q_{1D}Q_{2D}~,
\end{equation}
where $Q_{1D}$ and $Q_{2D}$ denote the configurational partition functions in one and two dimensions, respectively, and are given by
\begin{equation}\label{e8}
Q_{2D}=\frac{1}{S^{N}}\int{dx^{N}}dy^{N}
\exp\left(-\beta\frac{N-1}{2N}\phi(x_,y)\right)
\end{equation}
and
\begin{equation}\label{e9}
Q_{1D}=\exp\left(-N\beta{u_{pw}}(z^{*})\right)~,
\end{equation}
where $z^{*}$ is that value of the coordinate $z$ that guarantees the mean value of the Boltzmann factor \cite{martinez 2007}.

Then the final expression for the partition function has the following form
\begin{equation}\label{e10}
Z_{N}=Z^{ideal}_{2D}Q^{N}_{2D}\left(\frac{\lambda_{w}
\sigma}{\lambda_{B}}\right)^{N}\exp\left(N\beta{u_{pw}}\right)~.
\end{equation}
Applying the standard relation $A=-kT\ln Z$, the Helmholtz free energy of the adsorbed fluid is given by
\begin{equation}\label{e11}
\frac{A_{ads}}{NkT}=\frac{A_{2D}}{NkT}-\ln
\left(\frac{\lambda_{w}\sigma}{\lambda_{B}}\right)+\beta{u_{pw}}~,
\end{equation}
where $A_{2D}$ is the Helmholtz free energy of a two-dimensional fluid with interactions given by the binary potential
$\phi(x,y)$. In perturbation theory up to second order, the dimensionless Helmholtz free energy per particle, $A_{2D}/NkT$, can be written as follows
\begin{equation}\label{e12}
\frac{A_{2D}}{NkT}=\ln(\rho_{ads}\Lambda^{2})+
\frac{A_{HD}}{NkT}+\beta{a_{1}^{2D}}+\beta^{2}a_{2}^{2D}~.
\end{equation}
Here $A_{HD}$ is the Helmholtz free energy for hard disks, $\beta=1/kT$, $a_{1}^{2D}$ and
$a_{2}^{2D}$ are the first two terms of the $2D$ perturbation expansion. It is worth mentioning that the expression for $Q_{1D}$
as appears in Eq.~(\ref{e9}) is accurate and can be used to define the energy parameter of the wall, $\epsilon_{w}$. For the simple case where the wall-particle interaction is given by a square well with range $\lambda_{w}\sigma$ and energy depth $\epsilon_{w}$, we get
$u_{pw}(z^{*})=-\epsilon_{w}$.

A similar expression in perturbations is considered for the fluid in the bulk \cite{galindo 1998}
\begin{equation}\label{e13}
\frac{A_{3D}}{NkT}=ln(\rho_{b}\Lambda^{3})+ \frac{A_{HS}}{NkT}+\beta{a_{1}^{2D}}+\beta^{2}a_{2}^{2D}~.
\end{equation}
Getting the chemical potentials $\mu_{ads}$ and $\mu_{b}$ from
Eqs.~(\ref{e11}), (\ref{e12}), and (\ref{e13}), we can rewrite the condition (\ref{e5}) as follows
\begin{equation}\label{e14}
\mu_{3D}=\mu_{2D}+\mu_{w}~,
\end{equation}
where $\mu_{3D}$ and $\mu_{2D}$ are the chemical potentials in $3D$ and $2D$, respectively, and $\mu_{w}$ is the
wall contribution to the chemical potential
\begin{equation}\label{e15}
\mu_{w}=-\frac{1}{\beta}\ln(\lambda_{w})+{u_{pw}}(z^{*})~.
\end{equation}

\subsection{Adsorption of chain molecules}\label{sub2}
In the SAFT-VR framework the theory presented in Subsection \ref{sub1} can be easily extended
to the more general case of adsorption of homonuclear and heteronuclear chain molecules \cite{galindo 1998}, which is one of the tasks of this work.
In the simpler case of homonuclear molecules, we consider the fluid as made up of $N$
molecules each composed of $m$ identical spherical units (monomers) of diameter $\sigma$. We assume that both the
particle-particle and particle-wall interactions are described by square-well potentials. The thermodynamic equilibrium
between the bulk phase and the adsorbed phase of the fluid is obtained by imposing the condition in Eq.~(\ref{e14}) of equal chemical potentials.
The latter quantities are derived from the SAFT-VR expressions for the Helmholtz free energies of chain molecules in 3D and 2D.\\

{\em SAFT-VR approach}. The Helmholtz free energy $A$ of chain molecular gases
is composed of three separate terms \cite{chapman 1989,chapman 1990}: one
refers to the ideal gas free energy, another to the
free energy of monomeric units and finally the term corresponding to the
formation of chain molecules incorporating a given number of monomeric units
\begin{equation}\label{e16}
\frac{A}{NkT}=\frac{A^{ideal}}{NkT}+\frac{A^{mono}}{NkT}+
\frac{A^{chain}}{NkT}~.
\end{equation}
We will now provide the main expressions of each term in this equation in 3D and 2D.
For more details the reader is directed to refs.~\cite{martinez 2007,castro 2009,gil 1997}.

The ideal contribution is given by:
\begin{equation}\label{e17}
\frac{A^{ideal}_{3D}}{NkT}=\ln(\rho_{b}\Lambda^{3})~,
\end{equation}
\begin{equation}\label{e18}
\frac{A^{ideal}_{2D}}{NkT}=\ln(\rho_{abs}\Lambda^{2})~,
\end{equation}
where $\Lambda$ is the de Broglie thermal wavelength.
The monomer contribution is given by:
\begin{equation}\label{e19}
\frac{A^{mono}_{3D}}{NkT}=m\frac{A_{3D}}{N_{s}kT}~,
\end{equation}
\begin{equation}\label{e20}
\frac{A^{mono}_{2D}}{NkT}=m\frac{A_{2D}}{N_{s}kT}~,
\end{equation}
where $N_s$ is the number of monomer units of the chain molecule, and $A_{3D}$ and $A_{2D}$ are the free energy of monomer fluids in 3D and 2D, respectively, that can be obtained from perturbation theory according to Eqs.~(\ref{e11}) and (\ref{e13}).

Finally, the chain contribution is given by:
%
\begin{equation}\label{e21}
\frac{A^{chain}_{3D}}{NkT}=-(m-1)\ln y^{3D,SW}(\sigma)~,
\end{equation}
\begin{equation}\label{e22}
\frac{A^{chain}_{2D}}{NkT}=-(m-1)\ln y^{2D,SW}(\sigma)~,
\end{equation}
where $y^{3D,SW}$ and $y^{2D,SW}$ are the correlation functions in the 3D and 2D backgrounds obtained from
the corresponding radial distribution functions through the well known relationship $y(r)= g(r)e^{\beta{u(r )}}$.
Following the SAFT-VR procedure, the functions $g(r)$ are obtained
as perturbation expansions
\begin{equation}\label{e23}
g^{3D,SW}(\sigma)=g^{HS}(\sigma)+\beta\epsilon g^{3D}_{1}(\sigma)~,
\end{equation}
\begin{equation}\label{e24}
g^{2D,SW}(\sigma)=g^{HD}(\sigma)+\beta\epsilon g^{2D}_{1}(\sigma)~,
\end{equation}
where $g^{HS}$ and $g^{HD}$ are the radial distribution functions of the hard spheres and hard disks, respectively.

%
%

%
%
\section{Results}

\begin{figure} [x]
\centering
      \includegraphics[height=5.3cm]{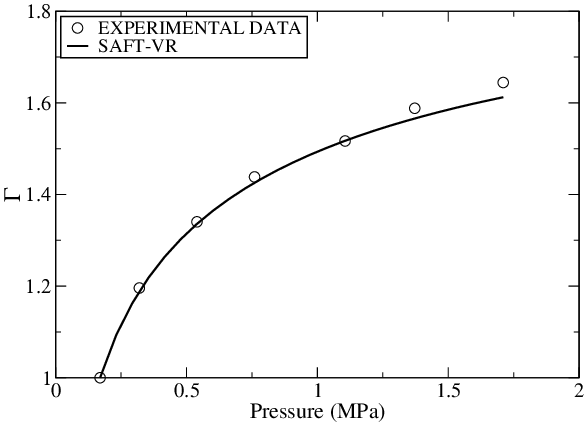}
  \caption{ Adsorption of $C_2H_4$ on activated carbon at 301.4 K. The circles are experimental data
from ref. \cite{relch 1980}.}\label{1etano}
\end{figure}

\begin{figure}
\centering
      \includegraphics[height=5.3cm]{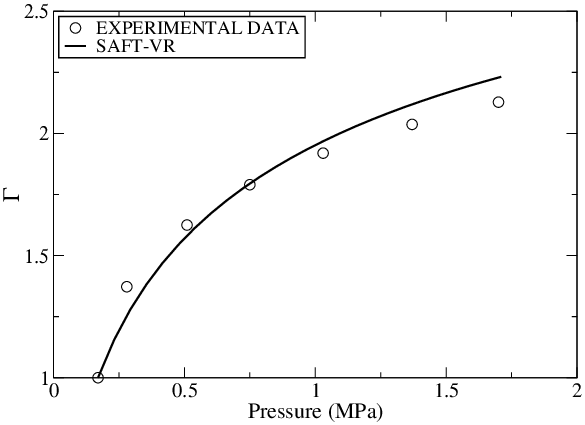}
  \caption{Same as in Fig.~\ref{1etano}, but for $C_2H_6$.}
  \label{2etileno}
\end{figure}

\begin{figure}
\centering
      \includegraphics[height=5.3cm]{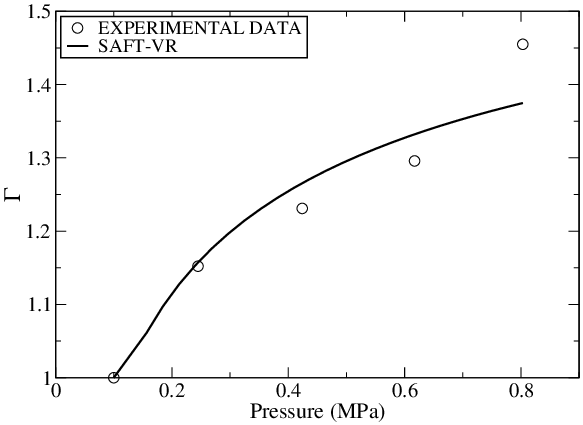}
  \caption{ Adsorption of $C_3H_8$ on activated carbon at 338.7 K. The circles are experimental data
from ref. \cite{ray 1950}.}\label{3propano}
\end{figure}

\begin{figure}
\centering
      \includegraphics[height=5.3cm]{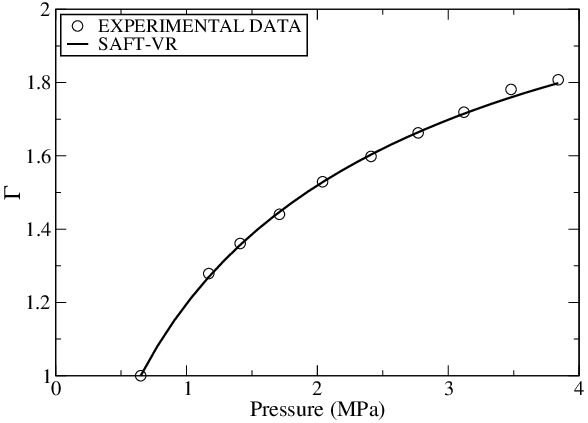}
  \caption{Same as in Fig.~\ref{3propano}, but for $CO_2$.}
  \label{4dioxido}
\end{figure}

\begin{figure}
\centering
      \includegraphics[height=5.3cm]{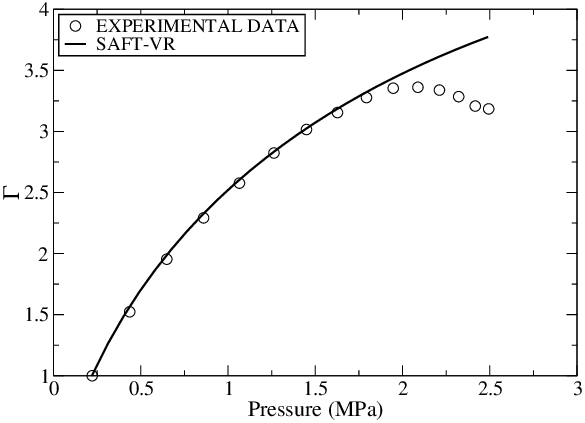}
  \caption{ Adsorption of $C_2H_4$ on silica gel at 278 K. The circles are experimental data
from ref. \cite{relch 1980}}\label{5etano}
\end{figure}

\begin{figure}
\centering
      \includegraphics[height=5.3cm]{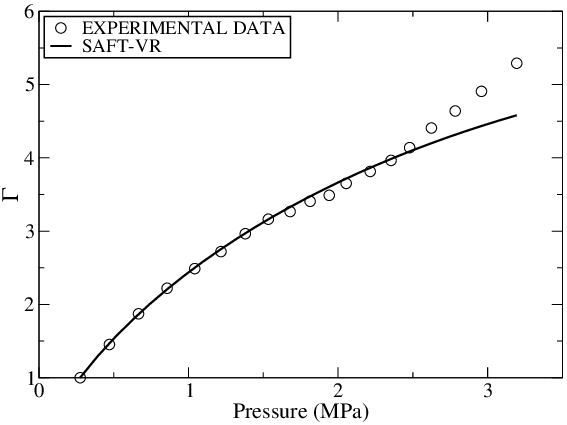}
  \caption{ Adsorption of $C_2H_4$ on silica gel at 303 K. The circles are experimental data
from ref. \cite{marie 1997}.}\label{6etano}
\end{figure}

\begin{figure}
\centering
      \includegraphics[height=5.3cm]{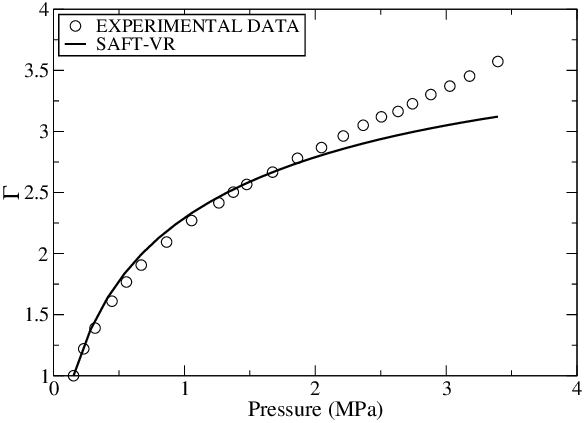}
  \caption{ Adsorption of $C_2H_6$ on silica gel at 278 K. The circles are experimental data
from ref. \cite{marie 1997}.}\label{7etileno}
\end{figure}

\begin{figure}
\centering
      \includegraphics[height=5.3cm]{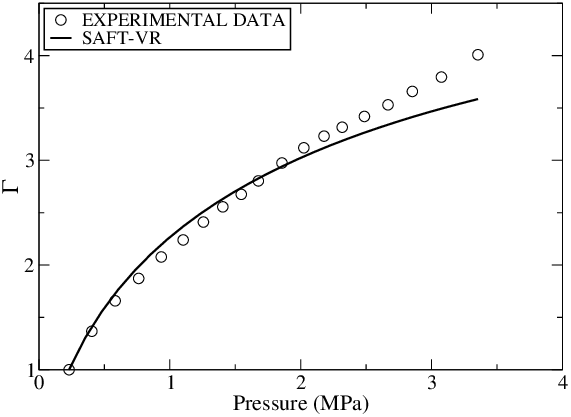}
  \caption{Same as in Fig.~\ref{7etileno}, but for $T=303$ K.}
  \label{8etileno}
\end{figure}

\begin{figure}
\centering
      \includegraphics[height=5.3cm]{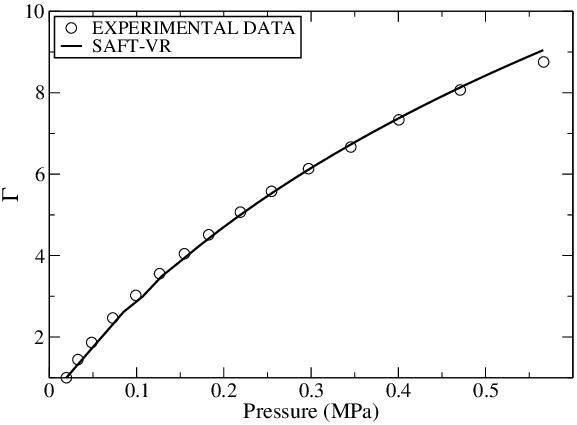}
  \caption{Adsorption of $C_3H_8$ on silica gel at 293 K. The circles are experimental data
from ref. \cite{marie 1997}.}\label{9propano}
\end{figure}

\begin{figure}
\centering
      \includegraphics[height=5.3cm]{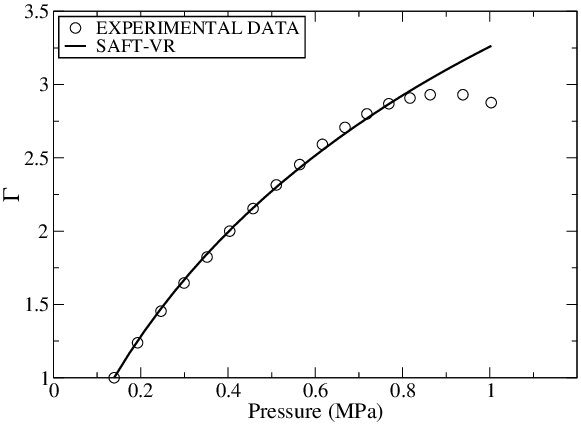}
  \caption{Same as in Fig.~\ref{9propano}, but for $T=303$ K.}
  \label{10propano}
\end{figure}

\begin{figure}
\centering
      \includegraphics[height=5.3cm]{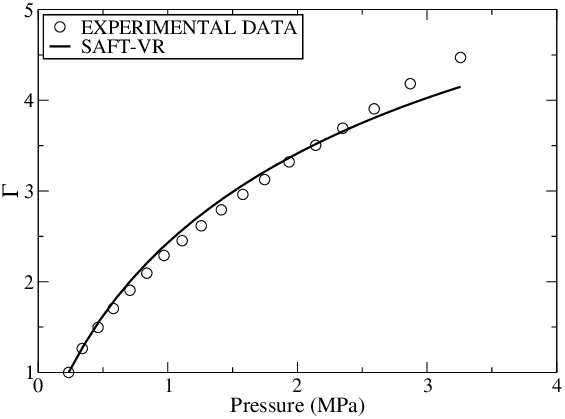}
  \caption{Adsorption of $CO_{2}$
on silica gel at 293 K. The circles are experimental data
see ref. \cite{marie 1997}.}\label{11co2}
\end{figure}

\begin{figure}
\centering
      \includegraphics[height=5.3cm]{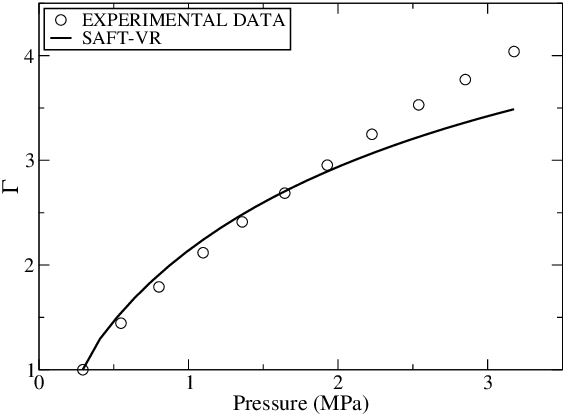}
  \caption{Same as in Fig.~\ref{11co2}, but for $T=303$ K.}
  \label{12dioxido}
\end{figure}

A comparison of the theoretical adsorption isotherms with the experimental results for the following gases
$C_2H_{4}$, $C_2H_{6}$, $C_3H_{8}$, and $CO_{2}$ onto activated carbon and silica gel is displayed in
Figs.~\ref{1etano}-\ref{12dioxido}
at different temperatures and pressures.
Table I and II show the parameters we have used to describe the investigated simple fluids
both in the bulk phase and the adsorbed phase within the SAFT-VR approximation \cite{martinez 2007,sagrario 2008,mc 1998,mcc 2001}.
As seen from these tables, ethane, ethylene, and propane are represented as
fluids of non-spherical monomers $(m=1.2, 1.33, 1.67)$, while $CO_{2}$ is represented as a
dimer fluid $(m = 2)$. The molecular diameters $\sigma$ are the same both in the bulk and the adsorbed phases, while the parameters of
the $SW$ attractive potential are different from case to case. These differences are due to
 the influence of the wall on the substrate of particles. As mentioned in previous works \cite{martinez 2007, castro 2009, sagrario 2008},
according to the theoretical results derived by Sinanoglu and Pitzer for a Lennard-Jones fluid \cite{pitzer 1960}, the energy of the
potential well depth for the particle-particle interaction in the adsorbed monolayer is reduced by a factor of $20$\%-$40$\% with
respect to their bulk phase values. We chose a reduction of $20$\% in all cases.
Once $\epsilon_{ads}$ is selected in this way, the range $\lambda_{ads}$ can be determined
by reproducing the experimental ratios between the critical temperatures of the bulk phase, $T_{c}^{b}$, and the adsorbed phase, $T_{c}^{ads}$, i.e.,
$R_{c}=T_{c}^{ads}/T_{c}^{b}$. The value of this ratio is known for the case of noble gases and methane adsorbed on graphite surfaces,
$R_{c}\approx{0.4}$. Next, $\lambda_{ads}$ was determined from this ratio and
the values of the set of parameters $m, \sigma, \epsilon, \lambda, \epsilon_{ads}$ for ethane, ethylene, propane and carbon
dioxide are those reported in Table 1 and 2. We notice here that at higher confinement regimes for the fluid, $R_c$ could depend on the degree of confinement for some types of activated carbon as well as for silica gels and therefore, strictly speaking, our results apply in the low confinement regime.

\begin{table}
\caption{The values of molecular parameters used to
describe the adsorption of ethane, ethylene, propane
and $CO_{2}$ on activated carbon.}
\vspace{5mm}
\begin{tabular}{lccccccccc}
\hline\hline
Substance&$m$&$\sigma(\AA)$&$\lambda$&$\epsilon/k$(K)&$\lambda_{ads}$&$
\epsilon_{ads}/k$(K)&$\lambda_{w}$&$\epsilon_{w}/\epsilon$&$T$(K)\\
\hline
$C_{2}H_{4}$&1.2&3.743&1.444&236.1&1.1592&188.88&0.2453&7.80&301.4\\
$C_{2}H_{6}$&1.33&4.233&1.449&224.8&1.1592&179.84&0.2453&9.8&301.4\\
$C_{3}H_{8}$&1.67&3.8899&1.4537&260.91&1.16296&208.728&0.8165&10.0&338.7\\
$CO_{2}$&2.0&2.7864&1.5257&179.27&1.2620&143.416&0.8165&7.7&301.4\\
\hline\hline
\end{tabular}
\end{table}
\begin{table}
\caption{The values of molecular parameters used to
describe the adsorption of ethane, ethylene, propane
and $CO_{2}$ on silica gel.}
\vspace{5mm}
\begin{tabular}{lccccccccc}
\hline\hline
Substance&$m$&$\sigma(\AA)$&$\lambda$&$\epsilon/k$(K)&$\lambda_{ads}$&$
\epsilon_{ads}/k$(K)&$\lambda_{w}$&$\epsilon_{w}/\epsilon$&$T$(K)\\
\hline
$C_{2}H_{4}$&1.2&3.743&1.444&236.1&1.1592&188.88&0.2453&4.5&278.0\\
$C_{2}H_{4}$&1.2&3.743&1.444&236.1&1.1592&188.88&0.2453&3.5&303.0\\
$C_{2}H_{6}$&1.33&4.233&1.449&224.8&1.1592&179.84&0.2453&6.3&278.0\\
$C_{2}H_{6}$&1.33&4.233&1.449&224.8&1.1592&179.84&0.2453&5.0&303.0\\
$C_{3}H_{8}$&1.67&3.8899&1.4537&260.91&1.16296&208.728&0.8165&2.69&293.0\\
$C_{3}H_{8}$&1.67&3.8899&1.4537&260.91&1.16296&208.728&0.8165&3.0&303.0\\
$CO_{2}$&2.0&2.7864&1.5257&179.27&1.2620&143.416&0.8165&4.95&293.0\\
$CO_{2}$&2.0&2.7864&1.5257&179.27&1.2620&143.416&0.8165&6.0&303.0\\
\hline\hline
\end{tabular}
\end{table}

For the range parameter $\lambda_{w}$ we used two values, $0.2453$ and $0.8165$. The
first one corresponds to the value found for the adsorption of
noble gases on graphite, whereas the latter one is the upper limit
used to describe the adsorption of a monolayer according to
the mean-field approach \cite{rio 1991}. The energy parameter was
adjusted to reproduce the experimental adsorption isotherms. In Figures \ref{1etano}-\ref{4dioxido}, the reported results for
the adsorption on activated carbon were conducted at temperatures
of 301.4 and 338.7 K at pressures up to 3.5 MPa.
One can notice the good agreement between theory and experimental data,
unless in Figure \ref{3propano} where our SAFT-VR calculations although not very close to the experimental data still follow their main trend.
Figures \ref{5etano}-\ref{12dioxido} show the results of adsorption on silica gel, for three different temperatures, i.e., 278, 293, and 303 K. We can see from these figures that the theoretical behavior agrees with the experimental data, except at high pressures where a deviation is noted which may
reflect the fact that the molecules are not spherical. Moreover, in the case of carbon dioxide the
quadrupolar contribution is not taken into account and it is known that it influences the thermodynamic properties of both the bulk phase and the adsorbed phases. Another feature that is not taken into account in the model is the way the molecules adhere to the wall,
that is, if their chain structure is perpendicular, parallel or
diagonal to the wall. Despite these limitations, the model provides very good theoretical results compared to the available experimental data.

An advantage of this SAFT-VR thermodynamic model is that one only needs to know the particle-particle and the wall-particle interaction energies together with the molecular diameter, and requires the adjustment of only one parameter that corresponds to the strength of the attraction energy of the wall, but this adjustment is based on the experimental data of the adsorbate energy $U_{w}$, which is known in the
literature as the isosteric heat of adsorption. For example, in the case of krypton adsorbed on graphite, $U_{w}$=11.72 kJ/mol \cite{ross 1964}
and corresponds to a value of $\epsilon$=8.168 \cite{rio 1991}, which is of comparable magnitude to the values used in this work.
On the contrary, other models use at least four
parameters to make adjustments and to reproduce the experimental data of the adsorption isotherms and usually these models are semi-empirical \cite{gasem 2003}.

\section{Conclusions}
We have applied the SAFT-VR approach to model adsorption isotherms of simple fluids with molecules of non spherical shape.
To achieve this goal we used a SAFT-VR approach in which the chemical equilibrium condition in both 3D
and 2D has been employed. For the cases of real fluids considered it is demonstrated that a good fit of these theoretical predictions is
obtained by adjusting only a single parameter, which is the energy of the attraction interaction of the molecules towards the wall.
We improved further the predictions by taking into account
the positions of the particles on the wall as well as adding a new term to the SAFT-VR-2D corresponding to the
quasi-two-dimensional quadrupole approximation. We recall that other semiempirical models available in the literature \cite{Marc,Mend} make use of at least four adjustable parameters without a clear physical interpretation, whereas the SAFT-VR parameters have a molecular basis being related to the molecular forces acting in the adsorption process.

On the other hand, the complicated porous structure of silica gels and some of the activated carbons is at the origin of the fractal character of their surfaces. Taking into account this fractality \cite{Nikl} and the distribution of pores in the SAFT-VR approach implies introducing fractal type parameters that remains as an interesting  issue for future investigation.

\section*{Acknowledgement}

The first author thanks CONACyT for a postdoctoral fellowship allowing him to work in IPICyT.
The refereeing suggestions are appreciated.

\end{document}